\newcommand{\vy}[2]{#1_{\scriptscriptstyle #2}}
\documentclass[12pt,referee]{aastex}

\shorttitle{Distance to multiple FeLoBAL components}
\shortauthors{Bautista et~al.}
\begin{document}
\title{Distance to Multiple Kinematic Components of 
Quasar Outflows: VLT Observations of QSO~2359--1241 and
SDSS~J0318--0600}

\author{Manuel A.\ Bautista\altaffilmark{1,2}, Jay P. Dunn,  Nahum Arav}
\affil{Department of Physics, Virginia Polytechnic and State University,
Blacksburg, VA 24061}

\author{Kirk T.\ Korista}
\affil{Department of Physics, Western Michigan University}
\affil{Kalamazoo, MI 49008-5252} 

\author{Maxwell Moe}
\affil{Department of Astronomy, University of Colorado, Boulder, CO}

\author{Chris Benn}
\affil{Isaac Newton Group, Observatorio del Rogue de los Muchachos, Spain}

\altaffiltext{1}{Currently at Department of Physics, Western Michigan University, Kalamazoo, MI 49008-5252}
\altaffiltext{2}{email: manuel.bautista@wmich.edu}

\begin{abstract}
Using high resolution VLT spectra, we study the multi-component
outflow systems of two quasars exhibiting intrinsic Fe II absorption
(QSO~2359--1241 and SDSS~J0318--0600). From the extracted ionic
column densities and using photoionization modeling we determine the
gas density, total column density, and ionization parameter for
several of the components. For each object the
largest column density component is also the densest, and all other
components have densities of roughly 1/4 of that of the main
component. We demonstrate that all the absorbers lie roughly
at the same distance from the source. Further, we calculate the total
kinetic luminosities and mass outflow rates of all components and show that these quantities are dominated by the main absorption component.
\end{abstract}

\section{Introduction}

 Rest-frame UV spectra of roughly 20\% of all quasars exhibit
blueshifted Broad Absorption Lines (BAL) that are indicative of an
outflow.  BAL are mainly associated with resonance lines of high
ionization species, like \ion{C}{4}, \ion{N}{5}, \ion{O}{6} (HiBAL),
and can reach velocities as high as 50,000 km $s^{-1}$ \citep{weymann, turnshek}.  
Despite various recent statistical studies of BAL quasars 
\citep{hall, trump, ganguly} the relationship between HiBAL and the host
galaxy remain illusive as these outflows do not contain distance
diagnostics in their spectra.  Thus, it is difficult to establish
whether the outflows affect only the near AGN environment (0.1-10~pc)
or whether they extend to the scales of the entire galaxy (1-10~kpc). 

A subset of BALQSOs also show absorption features from low ionization
species such as  \ion{Mg}{2}, \ion{Al}{2}, and most importantly for diagnostics, \ion{Fe}{2} and \ion{Si}{2}. These 
absorption features are generally
complex, as they are made of multiple components of narrower 
(of the order of a few hundred km~s$^{-1}$)
absorption troughs. AGN with this kind of features are often refer to as
FeLoBAL. In these systems the spectra of \ion{Fe}{2} and \ion{Si}{2}
is valuable because they  
often include absorption troughs from
metastable levels, which serve to determine 
the distance of the outflow from the central source and thereby beginning
to relate the outflows to their host galaxy. Such outflows were
studied in the past \citep{wampler, dekool01, dekool02,
hamann, arav01}. However, the presently available 
Sloan Digital Sky Survey (SDSS) 
and the development of advanced BAL analysis tools (Arav et al 2002,
Gabel 2005, Arav et al 2005) allow for a more accurate and systematic
study of these systems. With this in mind, we began a
comprehensive study of FeLoBAL outflows that contain distance diagnostics
\citep{arav08, korista08,
dunn09, moe09}.

Up to the present, though, all studies are based on either global 
properties of the multi-component absorption troughs or the strongest
kinematic component. 
Little is known about the individual 
physical properties of the components and their relationship to the
whole outflow.
In the best studied cases, \cite{wampler, dekool01, dekool02}
were able to obtain some column densities for individual components, but
could neither diagnose the physical conditions of the components nor derive
their relative distances. 
Simple assumptions that all components had similar physical conditions
and that they have either constant speed or equal monotonic acceleration  
would lead to the conclusion that either the absorbers are scattered along a wide 
range of distances from the central source or the absorbers vary in age as a result of 
episodic ejections. But BAL systems in general are far from simple. Apart from a weak 
correlations between the absorber bulk velocities and bolometric luminosity of the  
AGN
\citep{laor, dunn08, ganguly}, no correlations have been found among velocity, width, ionization, or 
any other observable properties. 

Recent high spectral resolution eschell observations of two quasars QSO 2359--1241 
and SDSS J0318--0600 with the Very Large Telescope (VLT) have allowed us to study 
in detail the various independent components that compose their FeLoBAL.
QSO 2359--1241 (VNSS J235953--124148) is a luminous ($M_B= -28.7$), reddened
($A_V \approx 0.5$) quasar at redshift $z=0.868$ \citep{brotherton01}. SDSS 
J0318--0600 is also a highly reddened bright quasar ($A_V\approx 1$ and $M_B=-28.15$) at redshift 
$z=1.967$. Both quasars exhibit a rich multi-component FeLoBAL comprising five and 
eleven components respectively. The strongest components in the FeLoBAL, which are 
also the first absorbers from the central engine (see Section 3), for each of these objects were measured 
and analyzed in \cite{arav08} and \cite{korista08} for QSO 2359--1241  and \cite{dunn09} for 
SDSS J0318--0600. It was found that the absorbers are located 
at $\sim$3~kpc and between 6 and 20 kpc 
from the central engine, respectively. 

QSO 2359--1241  and SDSS J0318--0600 are the first to be studied in detail from 
 a sample of $\sim$80 
FeLoBAL quasars with resonance and metastable \ion{Fe}{2} absorption lines (FeLoBAL 
quasars) in their spectra. These lines can be used for direct determination of the physical 
conditions and energy transport of the outflows.  The sample was extracted out of 
50,000 objects in the SDSS database as part of a major ongoing effort to study the 
nature of quasar outflows and their effects on the host galaxy \citep{moe09}.

\section{The measured column densities}
 
\begin{figure}
\rotatebox{0}{\resizebox{\hsize}{\hsize}
{\plotone{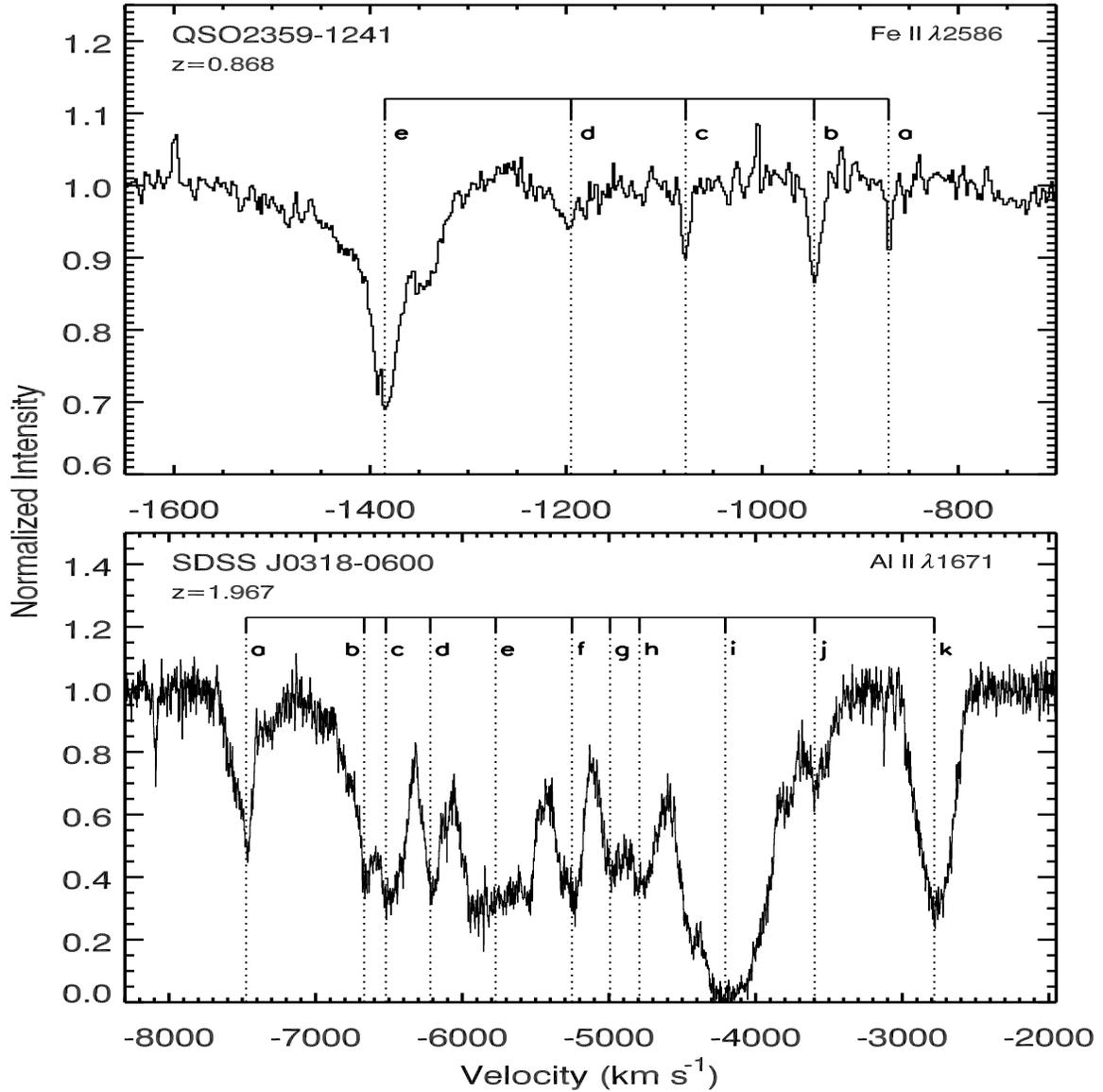}}}
\caption{Absorption troughs showing 5 and 11 clearly separated absorption
components in the spectra of QSO~2359--1241 and SDSS J0318-0600, respectively.
The same components in velocity space seem to be present in all troughs of
the same object. } 
\label{spectra}
\end{figure}

Observations of QSO~2359--1241 and SDSS J0318--0600 consist of 
echelle VLT/UVES high-resolution
($R \approx 40,000$) spectra with 6.3 hour exposures for each object. 
Fig.~\ref{spectra} illustrates the structure of the absorption troughs. The
observations for QSO~2359--1241 were presented in \cite{arav08}, together with the
identification of 
all the absorption features associated with the outflow.
Column densities of only the strongest component ($e$) were measured
at that time. 
The observations of SDSS J0318-0600 are presented in \cite{dunn09}. 
The high signal-to-noise data allowed us to measure the column densities 
from 5 unblended absorbers in QSO~2359--1241 
and 11 components in SDSS J0318--0600. For the latter, though, only the strongest 
components {\bf a, i}, and {\bf k} could be independently measured, while all other 
components had to be measured as a single blended structure. 
Whenever possible, the column densities were determined through three different 
assumption, i.e. full covering or apparent optical depth, partial line-of-sight 
covering, and velocity dependent covering according to the power law method of 
de Kool et al. (2001); Arav et al. (2005). For each component in 
QSO~2359$-$1241, we find that the troughs require the power law method to 
determine the full column density. We quote the column density determined by 
the power law method in Table 1. For the three components in SDSS~J0318$-$0600, 
we find that the results of these three methods are generally in good agreement,
indicating that there is full covering of each source and the column 
densities could be reliably measured (see Dunn et al. 2009). 

The measured column densities for the observed components in QSO~2359--1241
and SDSS J0318-0600 are given in Tables 1 and 2 respectively.

\clearpage

\clearpage 
\begin{deluxetable}{lcccccc}
\tabletypesize{\scriptsize}
\tablecaption{Measured Column Densities in QSO~2359--1241}
\tablewidth{0pt}
\tablehead{
\colhead{Species} &
\colhead{E$_{low}$ (cm$^{-1}$)} &
 \multicolumn{5}{c}{Column density ($\times 10^{12}$ cm$^{-2}$)}\\
 & &
\colhead{Comp. {\bf a}} &
\colhead{Comp. {\bf b}} &
\colhead{Comp. {\bf c}} &
\colhead{Comp. {\bf d}} &
\colhead{Comp. {\bf e}} 
}
\startdata
\ion{He}{1}& 159~856     & $14.3\pm0.7$ & $22.9\pm3.4$  & $6.1\pm0.3$  & $4.5\pm2.2$  & $138.0\pm9.9$  \\
\ion{Fe}{2} & 0    & $0.49\pm0.08$&  $5.7\pm1.5$  & $2.7\pm0.3$  & $2.7\pm0.1$  & $72.4\pm3.5$  \\
\ion{Fe}{2} & 385  &              & $0.93\pm0.28$ & $0.40\pm0.09$& $0.60\pm0.22$& $32.4\pm1.2$ \\
\ion{Fe}{2} & 668  &              & $1.1\pm0.5$   &              &         & $18.2\pm1.3$ \\
\ion{Fe}{2} & 863  &              & $0.45\pm0.20$ &              &         & $11.5\pm0.9$  \\
\ion{Fe}{2} & 977  &              &               &              &         & $7.1\pm0.6$  \\
\ion{Fe}{2} & 1873 &              &               &              &         & $77.6\pm9.5$  \\
\ion{Fe}{2} & 7955 &              &               &              &         & $5.0\pm0.5$  \\
\ion{Mg}{1} & 0    & $0.04\pm0.02$& $2.1\pm1.0$   & $0.28\pm0.02$& $0.04\pm0.03$& $0.83\pm0.06$ \\
\ion{Mg}{2} & 0    &              &               &              &         & $>65$  \\
\ion{Si}{2} & 0    &              & $198\pm48$    &              &         &  \\
\ion{Si}{2} & 287  &              &               &              &         & $794\pm206$ \\
\ion{Al}{3} & 0    &              &               & $12.7\pm0.4$ &         & $>79$ \\
\ion{Ca}{2} & 0    & $0.07\pm0.03$& $0.83\pm0.04$ & $0.47\pm0.02$&         & $3.3\pm0.3$ \\
\ion{Ni}{2} & 8394 &              &               &              &         & $6.2\pm0.6$ \\
\enddata
%\tablenotetext{a}{Upper limit based on the noise measurement method}
%\tablenotetext{b}{Estimated with a wall fitting technique using the Al II template}
%\tablenotetext{c}{Unreliable measurements}
%\tablenotetext{d}{Estimated using the tip fitting method}
\end{deluxetable} 

\begin{deluxetable}{lccccc}
\tabletypesize{\scriptsize}
\tablecaption{Measured Column Densities in SDSS J0318--0600}
\tablehead{
\colhead{Species} &
\colhead{E$_{low}$(cm$^{-1}$)} &
 \multicolumn{4}{c}{Column density ($\times 10^{12}$ cm$^{-2}$)}\\
 & &
\colhead{Comp. {\bf a}} &
\colhead{Comp. {\bf i}} &
\colhead{Comp. {\bf k}} &
\colhead{Comp. {\bf b-h}} 
}
\startdata
\ion{Al}{2} & 0   &$116.1\pm0.1$ & $400\pm40$    & $35.0\pm0.3$ & $>390$  \\
\ion{Al}{3} & 0   &$46.0\pm0.4$ & $1560\pm220$  & $73.0\pm0.8$ & $>810$  \\
\ion{C}{2}  & 0   &$333\pm11$   &               & $1100\pm200$ & $>14000$\\
\ion{C}{2}  & 63  &$577\pm21$   & $>19000$      & $1800\pm300$ & \\
\ion{C}{4}  & 0   &$734\pm10$   & $29000\pm3000$& $1297\pm13$ & $>10000$ \\
\ion{Fe}{2} & 0   &$<40$        & $1275\pm35$   & $154\pm6$ & $>490$\\
\ion{Fe}{2} & 385 &             & $294\pm77$    &            &  \\
\ion{Fe}{2} & 668 &             &               & $9.0\pm0.3$&\\
\ion{Fe}{2} & 863 &             & $147\pm36$    & &  \\

\ion{Fe}{2} & 977 &             &               & &  \\
\ion{Fe}{2} & 1873&             & $163\pm49$    & &  \\
\ion{Fe}{2} & 2430&             & $25.0\pm5.4$  & &  \\
\ion{Fe}{2} & 7955&             & $8.1\pm0.2$   & &  \\
\ion{Mg}{2} & 0   & $28.7\pm0.1$& $3200\pm400$  & $192\pm1$ & $>880$\\
\ion{Mn}{2} & 0   &             & $17.5\pm0.1$  & &   \\
\ion{Ni}{2} & 0   &             & $180 \pm4  $  & $<120$ & \\
\ion{Ni}{2} & 8394&             & $64.0\pm0.4$  & $10.0\pm0.3$ &\\
\ion{Si}{2} & 0   & $101\pm3$   & $7220\pm100$  & $640\pm150$ & $>3500$\\
\ion{Si}{2} & 287 & $<50$       & $7380\pm130$  & $352\pm12$ & \\
\ion{Si}{4} & 0   & $145\pm6$   & $5600\pm1300$ & $140\pm4$ & $>1800$\\
\enddata
\end{deluxetable}
\clearpage 

\clearpage
\begin{figure}
\rotatebox{0}{\resizebox{\hsize}{\hsize}
{\plotone{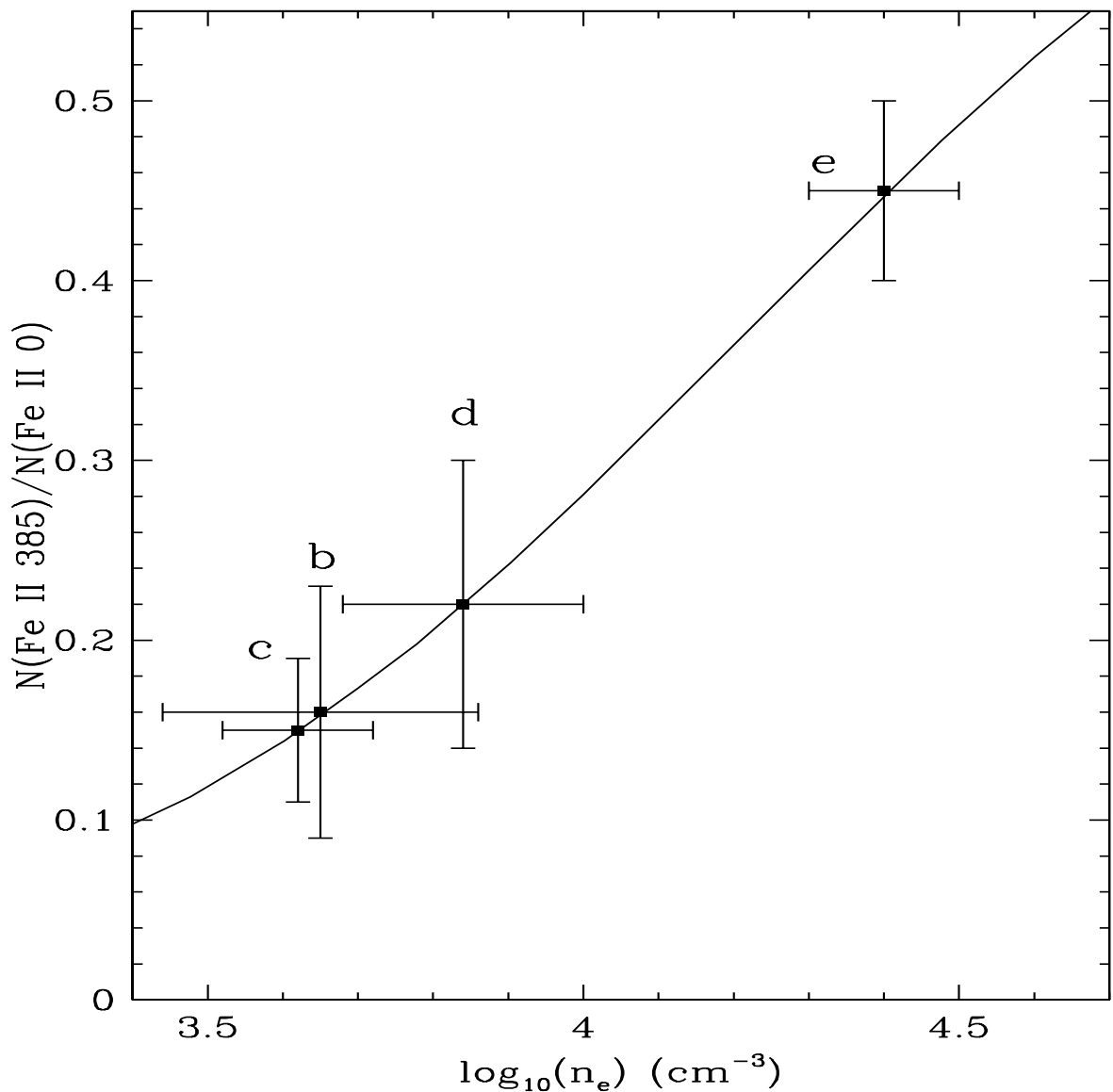}}}
\caption{Electron density diagnostics from \ion{Fe}{2} in QSO~2359--1241.
The ratio of column densities of
the excited level at 385~cm$^{-1}$ (a $^6$D$_{7/2}$) to the ground level 
(a $^6$D$_{9/2}$) is plotted against 
the logarithm of the electron density. The measured ratios for each of the
FeLoBAL components is drawn on top of the theoretical prediction. The
uncertainties in the measured ratios are depicted by vertical bars, and
that leads to uncertainties in the derived electron density which are 
indicated by horizontal bars.}
\label{fe2qso2359}
\end{figure}
\clearpage

\clearpage
\begin{figure}
\rotatebox{0}{\resizebox{\hsize}{\hsize}
{\plotone{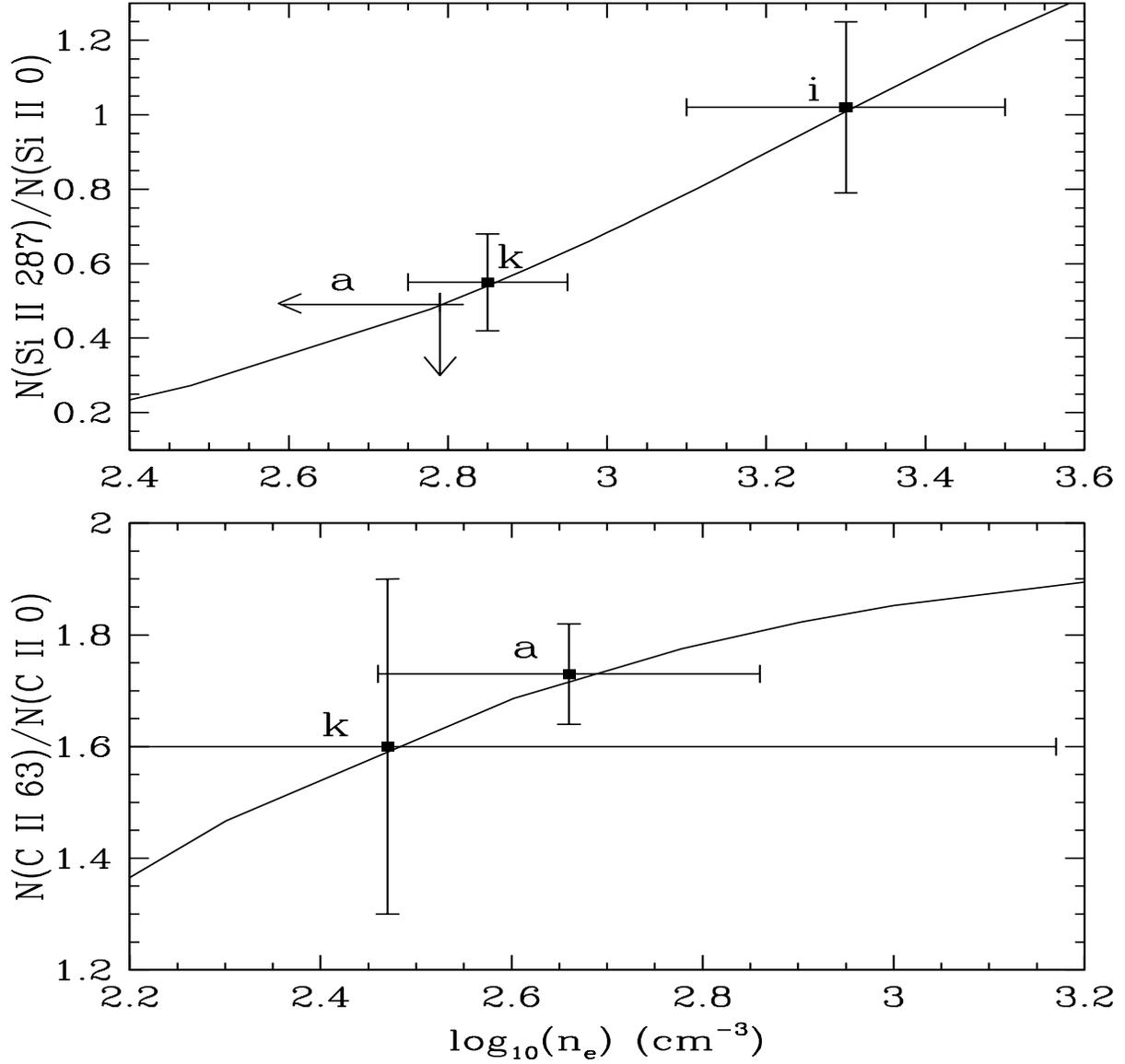}}}
\caption{Electron density diagnostics from the observed ratios of column densities of
excited to ground levels of \ion{C}{2} and \ion{Si}{2} in SDSS J0318-0600.
For the component {\bf a} only an upper limit to the \ion{Si}{2} ratio
could be obtained from observations.} 
\label{j0318}
\end{figure}
\clearpage

\section{Analysis of spectra and modeling}

\subsection{The density of the outflows}

We use the observed ratios of column densities of excited to resonance lines as electron 
density indicators, which are directly proportional to the level populations.
In  QSO~2359--1241 we find excellent diagnostics from the \ion{Fe}{2} column 
densities of the $a~^5D_{7/2}$ level at 385 cm$^{-1}$ and the ground level
$a~^5D_{9/2}$. The diagnostics are shown in Fig. \ref{fe2qso2359}. From these we 
get $\log(n_e/$cm$^{-3})=4.4\pm0.1$ for component {\bf e} (see Korista et al. 2008), 
$\log(n_e/$cm$^{-3})=3.8\pm0.2$ for component {\bf d},
and $\log(n_e/$cm$^{-3})=3.6\pm0.1$ and $3.6\pm0.2$ for components {\bf c} and {\bf b}.
Unfortunately,  we have no excited lines in component {\bf a} suitable for diagnostics.
The theoretical level populations for the \ion{Fe}{2} ion were computed from
the atomic model of \cite{baupra98}. 

In SDSS~J0318--0600 we find a density diagnostic for component {\bf a} in the ratio of 
\ion{C}{2} column densities of the excited level at 63 cm$^{-1}$ ($^2$P$^o_{3/2}$) to 
the ground level ($^2$P$^o_{1/2}$), which yields $\log(n_e/$cm$^{-3})=2.6\pm0.2$ cm$^{-3}$.
This is consistent with the limit $\log(n_e/$cm$^{-3})<2.8$ from the ratio of
the \ion{Si}{2} excited (287 cm$^{-1}$;  $^2$P$^o_{3/2}$) to the ground  
($^2$P$^o_{1/2}$) levels. 
For components {\bf i} and {\bf k} 
the \ion{Si}{2} ratios 
yield $\log(n_e/$cm$^{-3})=3.3\pm0.2$ 
and $\log(n_e$/cm$^{-3}$)=$2.85\pm0.10$ respectively. Additional diagnostics from 
\ion{Fe}{2} are available for component {\bf i} and they are all consistent with the present 
determination (see Dunn et al. 2009). Fig. \ref{j0318} illustrates the present  
diagnostics.
The theoretical level populations for \ion{C}{2} were computed using the
effective collision strengths of \cite{blum} and A-values from
\cite{wiese}. For the \ion{Si}{2} spectral model we use the
effective collision strengths of 
Dufton \& Kingston (1991) and 
A-values for forbidden transitions of
Nussbaumer (1977). 

In \cite{arav08}, \cite{korista08}, and \cite{dunn09} we showed that
the troughs in QSO~2359--1241  and SDSS~J0318--0600
arise from a region where
hydrogen is mostly ionized, thus the electron density derived above
should be nearly equal (within $\sim$10\%) to the total hydrogen
density of the clouds. 

 \subsection{Ionization structure of the outflows}

Under the premise that the absorbers are in photoionization equilibrium we 
assume constant gas density clouds in plane parallel geometry. Thus, 
the ionization structure of a warm photoionized plasma are typically characterized by 
the so called ionization parameter, which is defined as
\begin{equation}
U_H \equiv {\Phi_H\over{n_H c}} = {Q_H\over {4\pi R^2 n_H c}}, 
\end{equation} 
where  $c$ is the
speed of light, $n_H$ is the gas density,  
$\Phi_H$ is the ionizing photon flux, $Q_H$ is the rate of hydrogen ionizing photons
emitted by the ionizing source, and $R$ is the distance from the ionizing source to the 
cloud. From this definition, $Q_H$ can be estimated directly from the 
luminosity of the object and some knowledge of its Spectral Energy 
Distribution (SED).

\cite{korista08} and \cite{dunn09} constructed detailed photoionization models of the main
components of QSO~2359--1241 and SDSS~J0318--0600 and determined
their physical conditions.
But, modeling all the absorbing components of the outflow simultaneously opens
a variety of new scenarios. Fortunately, the combined measurements of column
densities for \ion{Fe}{2} and \ion{He}{1} in all of the components of QSO~2359--1241
and \ion{Si}{2} and \ion{Si}{4} in the components of SDSS~J0318--0600  allow us to constrain the
ionization parameter of all of these components.

\begin{figure}
\rotatebox{0}{\resizebox{\hsize}{\hsize}
{\plotone{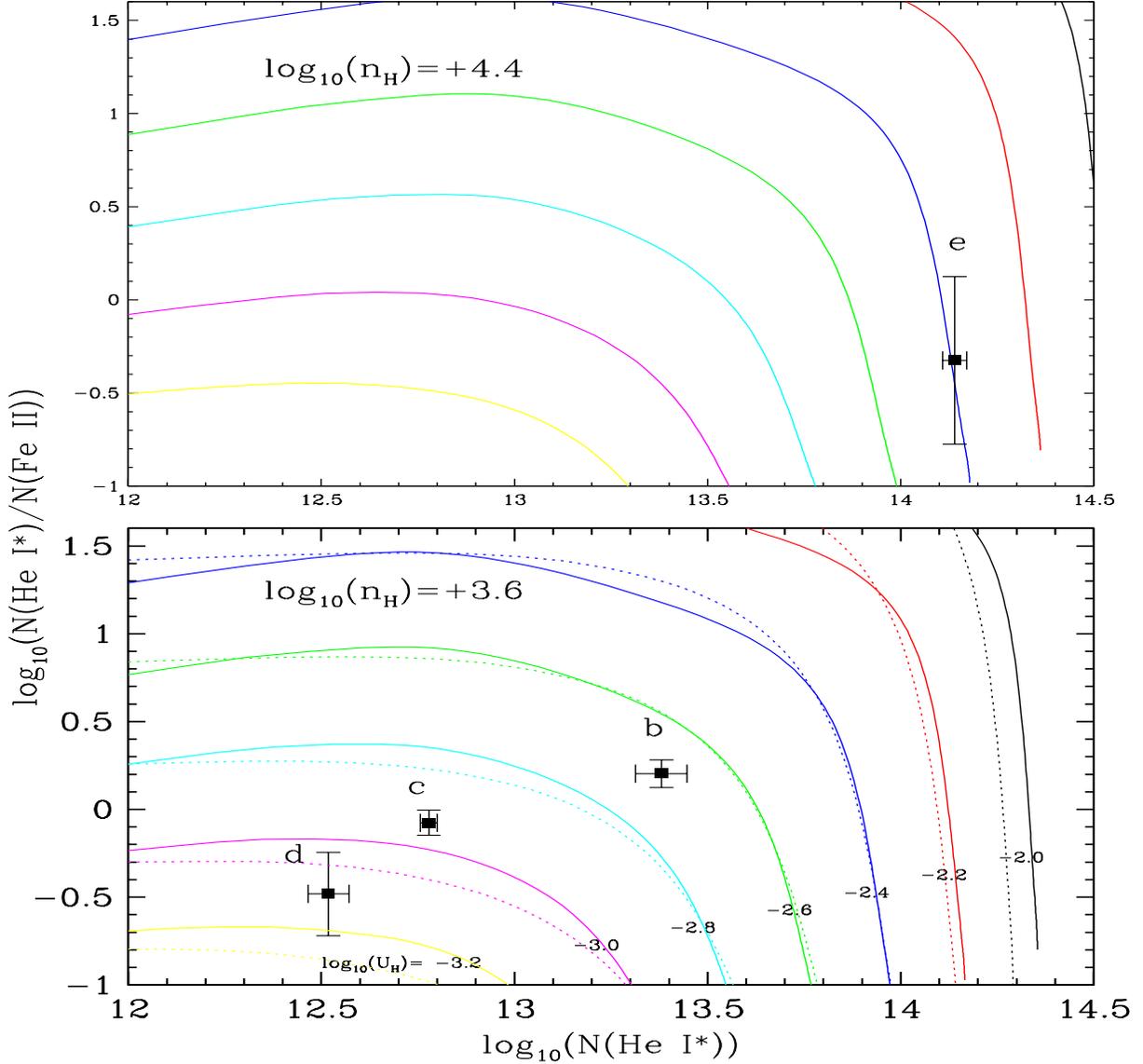}}}
\caption{Results of grid models for QSO~2359--1241 and diagnostics
of $U_H$ and total column density for all the kinematic 
absorption components of the outflow. The various curves presented are
for $\log_{10}(U_H)=-2.2$ (red), -2.4 (blue), -2.6 (green),
-2.8 (cyan), -3.0 (magenta), -3.2 (yellow). The solid lines depict the
results from the MF87 SED and the dotted lines show the results
from the spectrum transmitted through (attenuated by) component {\bf e}.
The calculated column densities are plot for three different
values of $\vy{n}{H}$, in agreement with the previous density diagnostics.}
\label{columnqso2359}
\end{figure}

We use the photoionization modeling code {\sc cloudy}~c07.02.01 \citep{cloudy} to compute grids 
of models in $U_H$ for each value of $\vy{n}{H}$ of interest. All 
models are calculated for $N_H$ (total hydrogen column density) running from the inner phase of the cloud to deep 
into the ionization fronts (IF's) where the temperature drops to only a few 
thousand K. For QSO~2359--1241  we assumed solar abundances as in 
\cite{korista08}. Then, the accumulated column densities of 
the metastable
2~$^3S$ excited state of \ion{He}{1} (hereafter, \ion{He}{1}$^*$)
and total 
\ion{Fe}{2} were extracted out of the {\sc cloudy} output files and 
plotted in Fig.~\ref{columnqso2359} as 
N(\ion{He}{1}*)/N(\ion{Fe}{2}) vs. $\log$N(\ion{He}{1}*). These plots serve as
direct diagnostics of $U_H$ by marking the measured column densities on the plots. 
Further, the value of $N_H$ is also readily available as there is a direct correspondence 
in each model between the accumulated total hydrogen column and
the accumulated column density of all species. 

The diagnostic power of the combined \ion{He}{1}$^*$ and \ion{Fe}{2} lines
was explained in \cite{korista08}. 
It was shown that the \ion{He}{1}$^*$ column density is set by the \ion{He}{2} column, which 
is bound to the \ion{H}{2} fraction and 
the ionization parameter. This is because whenever helium is ionized hydrogen, with a much lower ionization potential, must be ionized too. For any conceivable quasar ionizing continuum recombination of He II to neutral He occurs simultaneously with recombination of H II to neutral H. 
This is shown in Fig. 1 of Korista et al. (2008). On the other hand, the ionization fraction of
\ion{Fe}{2} across the ionization threshold is determined by charge exchange
with hydrogen
$${\rm Fe}^{++} + {\rm H}^0 \rightleftharpoons {\rm Fe}^{+} + {\rm H}^+.$$
Thus, the column density of \ion{Fe}{2} traces the column of neutral hydrogen
and
\begin{equation}
{N(He~I^*)\over N(Fe~II)} \propto 
{N(H~II)\over N(H~I)} \propto U_H;
\end{equation}
in other words, the measured ratio of column densities of \ion{He}{1}$^*$ 
to \ion{Fe}{2} serves as direct indicator of the average ionization parameter
of the gas cloud.
Within the cloud, whose depth, $r$, is expected to be much smaller than $R$,
the $U_H$ varies as 
\begin{equation}
U_H(r) \propto {e^{-\tau(r)}\over n_e}, 
\end{equation}
where $\tau$ is the optical depth to hydrogen ionizing photons, which 
is proportional to the column density of neutral hydrogen. 
Hence, at the inner phase of the cloud and for a large fraction of
it hydrogen is nearly fully ionized and $\tau$ remains small. Within
this depth in the cloud the ionization of the gas and hence the
$N$(\ion{He}{1}$^*$)/$N$(\ion{Fe}{2}) ratio is expected to remain 
roughly constant (see figure 8 in Arav et al. 2001). Deeper in the cloud, a small decrease in the ionization
of hydrogen leads to an increase in $\tau$, which in turns reduces
the number of ionizing photons right ahead 
and results in even lower hydrogen
ionization. At this point, an ionization front 
develops and is accompanied by a quick drop of the
$N$(\ion{He}{1}$^*$)/$N$(\ion{Fe}{2}) ratio into the cloud or,
equivalently, $N$(\ion{He}{1}$^*$).
Furthermore, the $N$(\ion{He}{1}$^*$)/$N$(\ion{Fe}{2}) vs. $N$(\ion{He}{1}$^*$)
plots shown in Fig.~3 offer a complete diagnostic for $U_H$ and
the total hydrogen column density ($N_H$) for clouds 
of a given chemical composition. 

Here, it is important to realize that as these plots are sensitive to
the integrated flux of hydrogen ionizing photons, most of which 
($\sim90 \%$) arise
from the spectral region between 1 and 4~Ry in Quasar SED. Thus,
the plots depend only little on the actual shape of the SED.
This is illustrated in Fig.~3 by showing the theoretical curves as 
calculated with the 
Mathews and
Ferland (1987; MF87 hereafter) quasar SED
 and with 
the transmitted spectrum from the inner most component ({\bf e}) of the outflow. 
That component {\bf e} is the first absorber from the source
in QSO~2359--1241
becomes apparent as this has the
largest $U_H$~dex and is also the densest, thus it is concluded from Eqn.~1 that it most have the smallest distance $R$ among all components (see also the discussion below).

From the diagnostics in Fig.~4  it results that component {\bf e} 
has $\log(U_H)$= -2.4 for $\log_{10}(n_H/$cm$^{-3})=4.4$. 
Component {\bf d} with a density $\log_{10}(n_H/$cm$^{-3})=3.8$ has
$\log_{10}(U_H)\approx -2.8$, and the two lowest density components {\bf b} and
{\bf c} have  $\log_{10}(U_H)\approx -2.7$ and -2.9 respectively.
{\bf Notice that the electron densities of components {\bf b} and 
{\bf d} are sufficiently similar to each other, and in fact overlap
within the uncertainty bars, for them to be plotted on the same diagnostic
diagram without any significant lose of accuracy.}

\begin{figure}
\rotatebox{-90}{\resizebox{\hsize}{\hsize}
{\plotone{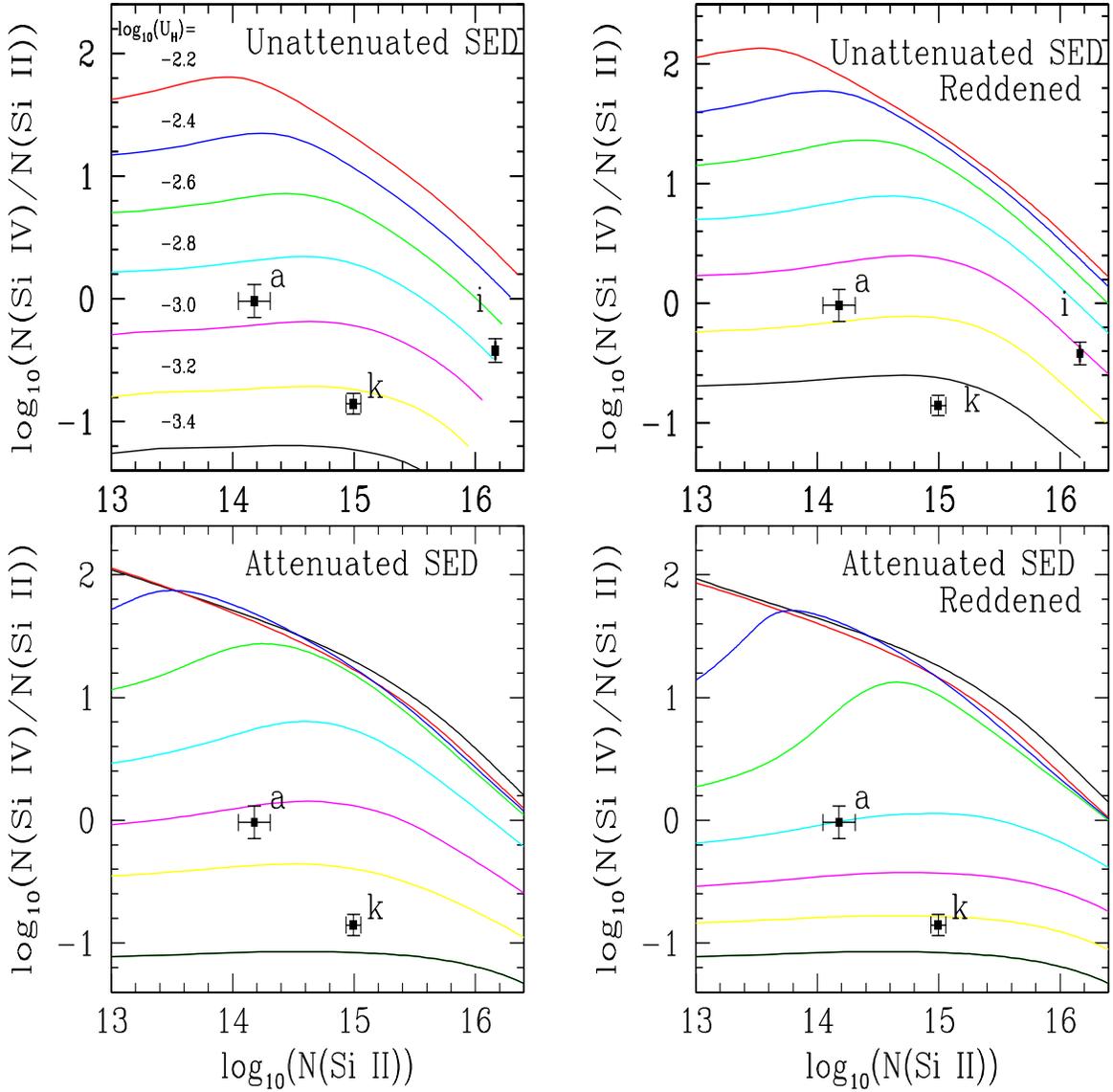}}}
\caption{Results of grid models for SDSS J0318--0600 and diagnostics
of ionization parameters and total column density for all kinematic 
absorption components of the outflow. The upper panels shows the results
with unattenuated unreddeneded SED (left) and reddeneded SED (right), while the lower panels presents the results from
the unreddened SED (left) and reddened SED (right) 
after attenuation by component {\bf i}. All models we used
$\log(\vy{n}{H})$=3.0, but are practically independent of 
$\vy{n}{H}$.
The different colors of the curves correspond to 
$\log_{10}(U_H)=-2.2$ (red), -2.4 (blue), -2.6 (green),
-2.8 (cyan), -3.0 (magenta), -3.2 (yellow), -3.4 (black).}
\label{columnj0318}
\end{figure}

We build similar plots for SDSS J0318--0600 but based on the column densities of
\ion{Si}{2} and \ion{Si}{4} (Fig~\ref{columnj0318}). For these models we start with a 
chemical composition expected for a galaxy with metalicity $Z$=4.2, which
we found to be a reasonably choice that fits well the measured column densities (Dunn et al. 2009). Models with either solar composition or very high metalicities
(e.g $Z=$7.2) can be discarded on the basis of the observed absorption
troughs for various species. Basing the diagnostics on two 
ions of the same element has two important advantages over the previous plots: (1) 
the column density ratio is mostly independent of the assumed chemical abundances, and 
(2) the plots are mostly independent of $\vy{n}{H}$. A disadvantage, 
though, is that the \ion{Si}{2} is mostly created through photoionization
by radiation below the hydrogen ionization threshold (0.6~Ry), while
\ion{Si}{4} has a formation energy higher than that of \ion{He}{2}.
Consequently, the $N$(\ion{Si}{4})/$N$(\ion{Si}{2})
is more sensitive to the SED than in the previous case.
Because SDSS J0318--0600 is a extremely reddened object and 
the location of the extinguishing dust with respect to the absorber
is unknown we need to consider two different SEDs for the models. These
are: the UV-soft SED developed in Dunn et al (2009) and this SED after 
reddening.
In Fig.~4 we plot  $\log(N$(\ion{Si}{4})/$N$(\ion{Si}{2}))
vs. $\log_{10}(N$(\ion{Si}{2}) as obtained from the two SEDs 
considered (upper panels) 
and in the cases in which these SEDs are attenuated by
component {\bf i}, 
which is identified as the inner most absorber (lower panels). 

From these diagnostics 
the densest component {\bf i} has the highest $U_H$ (=$-2.75\pm0.10$ dex
for the unreddened SED and $3.02\pm0.10$~dex for the reddened SED) 
and largest
total column (=$20.9\pm0.1$~dex for the unreddened SED and $20.1\pm0.1$~dex for the reddened SED). The lower density components {\bf a} and {\bf k} have 
considerably less column density and are less ionized. 

From the physical conditions derived above it is now possible to estimate the distance
from each absorption component to the ionizing source. But, it is convenient
to determine first the relative distances of all component to the source.
The relative distance can be obtained more accurately from
observations than absolute distance. This is because the absolute determination of distance depends strongly on the number of ionizing photons of the SED and on whether this is reddened before ionizing the cloud. So, we see that in 
QSO 2359--1241 and SDSS J0318--0600 the absolute distance to the absorbers vary by several factors depending on whether reddening of the SED occurs before or after ionizing the absorbers. This sort of uncertainty, however, does not affect the relative distances among absorbers. From Eqn.(1) one gets
\begin{equation}
\log(R/R_0) = {1\over 2} [\log(Q_H/Q_{H0}) - \log(\vy{n}{H})/\vy{n}{H0}) - \log(\vy{U}{H}/\vy{U}{H0}) ],
\end{equation}
where $R_0$, $\vy{Q}{H0}$, $\vy{n}{H0}$, and $\vy{U}{H0}$ are
the distance, rate of ionizing photons, particle density, and ionization
parameter for a given reference component, which we choose as the
strongest component in each of the absorbers studied (i.e. {\bf e} in
QSO 2359-1241 and {\bf i} in SDSS J0318--0600). In identifying the inner-most 
absorber it is important to realize that in the first approximation that uses 
the same SED in determining the distance to all components the result is 
actually correct for the inner-most component and overestimated for the rest 
(see Eqn. 4 of the manuscript). This means that the absorber with the shortest 
distance to the source in this first approximation is indeed the inner-most 
absorber. For any other absorber to located in the inner-most position the 
other absorbers with shorter distance in the first order approximation would 
have to be at larger distances that initially estimated, which is impossible 
if the flux of ionizing photons is to be reduced by the effect of attenuation. 
Notice that the relative locations of all the components could, in principle, 
be determined following the same logical argument, except that they are so 
close together that their estimated distance differences soon become smaller 
than the uncertainties.
The strongest
components are also the innermost, as we will see below.
For any absorption
component that sees the same unattenuated radiation from the source
as the reference component
$\vy{Q}{H}=\vy{Q}{H0}$. On the other hand, if a component is
shadowed by intervening gas, particularly by the reference component,
$\vy{Q}{H}<\vy{Q}{H0}$.

Table~\ref{distances} presents the relative distance for
components in the two absorbers considered. First, we assume that
all components see the same unattenuated source ($\vy{Q}{H}=\vy{Q}{H0}$)
and the results are
given in columns 4 and 5 of the table. Under this
assumption components {\bf e} and {\bf i} are the inner-most absorbers in QSO 2359--1241
and SDSS J0318--0600 respectively. Beyond these, other components are
dispersed along 2 to 4 times that distance. We note that there are no
correlations between velocity, $\vy{n}{H}$, and $R$.

\clearpage

\clearpage
\clearpage
 
\clearpage
%\nonumber
\begin{deluxetable}{ccrcccccrrr} 
\rotate
\tabletypesize{\scriptsize}
\tablecaption{Calculated distances to the outflows}
\tablewidth{0pt}
\tablehead{
\colhead{Comp.} & 
\colhead{$v$ (km/s)}& 
\colhead{$\delta v$ (km/s)\tablenotemark{a}} & 
\colhead{$\log(n_H)$}&
\multicolumn{2}{c}{Unattenuated SED}&
\multicolumn{2}{c}{Attenuated SED\tablenotemark{b}}
 & & & 
\cr
\colhead{}&
\colhead{}&
\colhead{}&
 &
\colhead{$\log(U_H)$}& 
\colhead{$\log(R/R_0)$}&
\colhead{$\log(U_H)$} &
\colhead{$\log(R/R_0)$}&
\colhead{$\log(N_H)$} &
\colhead{$\dot E/\dot E_0$} &
\colhead{$\dot M/\dot M_0$}
}
\startdata
\multicolumn{10}{c}{QSO 2359--1241}\cr
b& -945	& $14\pm5\ $ &$3.65\pm0.21$&$-2.68\pm0.05$&
$+0.5\pm0.2$&$-2.68\pm0.05$&
$+0.0\pm0.2$ &$20.04\pm0.07$& 0.10 & 0.21\cr
c&-1080	& $30\pm21$ &$3.6\pm0.1$&$-2.89\pm0.05$&
$+0.7\pm0.1$&$-2.87\pm0.05$&
$+0.2\pm0.1$ & $19.46\pm0.05$& 0.06 & 0.06\cr
d&-1200	& $28\pm12$ &$3.84\pm0.16$&$-2.83\pm0.05$&
$+0.5\pm0.2$&$-2.81\pm0.05$&
$+0.0\pm0.2$ & $19.66\pm0.05$& 0.08 & 0.11\cr
e&-1380	& $59\pm22$ &$4.4\pm0.1$&$-2.42\pm0.03$&
      &                & 
    & $20.56\pm0.05$ &  1 & 1\cr
\\
\hline
\multicolumn{10}{c}{SDSS J0318--0600, unreddened SED} \cr
a  &-7450 & $150\pm 10$ &$2.65\pm0.20$ &$-3.04\pm0.10$ &$+0.5\pm0.2$& 
$-3.02\pm0.05$ &
$-0.3\pm0.2$ & $18.2\pm0.2$ & 0.01 & 0.004\cr
i  &-4200 & $670\pm 10$ &$3.3\pm0.2$ &$-2.75\pm0.10$ &
      &              & 
    & $20.9\pm0.1$	 & 1 & 1\cr
k  &-2800 & $290\pm10$ &$2.85\pm0.2$ &$-3.30\pm0.10$&
$+0.5\pm0.2$&$-3.40\pm0.10$&
$-0.2\pm0.2$ & $18.8\pm0.3$& 0.002 & 0.005\cr
\hline
\multicolumn{10}{c}{SDSS J0318--0600, reddened SED} \cr
a  &-7450 & $150\pm 10$ &$2.65\pm0.20$ &$-3.13\pm0.10$ &$+0.4\pm0.2$&
$-2.80\pm0.05$ &
$-0.2\pm0.2$ & $18.2\pm0.2$ & 0.07 & 0.02\cr
i  &-4200 & $670\pm 10$ &$3.3\pm0.2$ &$-3.02\pm0.10$ &
      &              &
    & $20.1\pm0.1$       & 1 & 1\cr
k  &-2800 & $290\pm10$ &$2.85\pm0.2$ &$-3.40\pm0.10$&
$+0.4\pm0.2$&$-3.19\pm0.10$&
$-0.1\pm0.2$ & $18.8\pm0.3$& 0.01 & 0.03\cr
\hline
\enddata
\tablenotetext{a}{Full width half maximum measured from \ion{Fe}{2}
in QSO~2359--1241 and \ion{Al}{2} for SDSS J0318-0600} 
\tablenotetext{b}{The scales for QSO 2359--1241
and SDSS J0318-0600 are 
($R_0$, $\dot E_0$, 
$\dot M_0$) =  
($1.3\pm0.4$ kpc, $2\times 10^{43}$ ergs~s$^{-1}$, 40 M$_\odot$ yr$^{-1}$)
and
($6\pm 3$ kpc, $1\times 10^{45}$ ergs s$^{-1}$, 180 M$_\odot$ yr$^{-1}$) 
if rednig of the SED occurs before ionizing the outflow or 
($3\pm1$ kpc, $4\times 10^{43}$ ergs $^{-1}$, 100 M$_\odot$ yr$^{-1}$) 
and ($18\pm8$ kpc, $6\times 10^{45}$ ergs s$^{-1}$, 1100 M$_\odot$ yr$^{-1}$) 
if redning occurs after.} 
\label{distances}
\end{deluxetable}
\clearpage

However, given that the distance scales from the absorbers to the central source
(kpc scales) are much greater than the size of the central source
it seems much more physically plausible 
that as the innermost components will shadow all further absorbers.
Thus, as the first absorber is ionized by the source the next
component in line from the source will
 only receive the transmitted SED from the first absorber, 
i.e. an attenuated SED. Furthermore, every component would only
see SED attenuated by all absorbers closer to the source. Thus, in
Table 3 we recalculate the distance for all components, other than the inner most,
using SEDs that account for the attenuation by the inner most components.
In this case $\log(\vy{Q}{H}/\vy{Q}{H0})=-1.0$ for QSO2359--1241 and
-1.64 or -0.80 for SDSS J0318--0600 when using the unreddened and reddened
SED respectively. 
Surprisingly, all components of both objects converge, within the uncertainties,
to the same distance from the central source. The uncertainties 
quoted in the table combine the errors in the values of $\vy{n}{H}$ and $\vy{N}{H}$ as
diagnosed from the measured column densities from spectra.
In both quasars studied here the inner most absorbers are clearly identified
as the densest and largest systems. The relative ordering
of subsequent absorbers with respect to the central source could be 
tentatively 
estimated too, but we make no attempt to do so because 
the distance between them is always smaller than the uncertainties.

Also in Table~\ref{distances}  
we present for every component 
the estimated kinetic luminosity and mass flux rate, defined as
\begin{equation}
\dot E = 4\pi \mu m_p \Omega R N_H v^3,
\end{equation}
\begin{equation}
\dot M = 8\pi \mu m_p \Omega R N_H v,
\end{equation}
were $\mu\approx 1.4$ is the mean particle mass for solar composition, $m_p$ is the proton mass, and 
$\Omega$ is the global covering
factor. 
For the present calculations we adopt $\Omega=0.2$ (see section 4.2 in Dunn et al. 2009),
which impacts the absolute values $\dot E_0$ and $\dot M_0$ given in the
table but not the relative contributions of the components.
These quantities are clearly dominated by the contribution of the largest 
innermost component of each outflow, while the minor components together contribute 
little to the total energy and mass carried out by the outflow. 
This indicates that the minor 
components may not be considered as ejection events in their own merits,
but instead they are physically related to the main component. 

Clearly, if attenuation of the SED were ignored the distance
to each of the minor components would be overestimated and as well as
their $\dot E$ and $\dot M$ contributions. Yet, even in this case
all the minor components together could only account for a small fraction of 
$\dot E$ and $\dot M$ sustained by the main component. 

{\bf 
Finally, in Table~3 we quote the absolute values of $R_0$, $E_0$, and $\dot M_0$
for the main components of QSO 2350--1241
and SDSS J0318--0600
as obtained in Korista et al (2008) and Dunn et al. (2010). These absolute
value of $R_0$ is a lot more uncertain than the relative quantities tabulated here,
for the reasons explained already at the begining of h=this section. The determination of absolute kinetic
energies and mass outflow rates are even more uncertain because they depend
on $R$ and the assumed valued for the global covering factor, which is
least known parameter of the investigation.}

\section{Discussion and conclusions}

The high spatial resolution and signal-to-noise of the VLT spectra 
allowed us to study the properties of each of the kinematic
components in the FeLoBALS of quasars QSO 2350--1241 
and SDSS J0318--0600. 
From the measurements of column densities for different kinematic components
we determined the electron number density for these components.
For QSO 2350--1241 we used the ratio of column densities of   
\ion{Fe}{2} from the excited level at 385 cm$^{-1}$ and the ground
level. In the case of SDSS J0318--0600 we used the ratios of 
column densities among levels of the ground multiplets of \ion{Si}{2}
and \ion{C}{2}. Interestingly, there is a clear density contrast between
the maint kinematic component in each object and the smaller components. 
By contrast, all smaller components in each object seems to exhibit roughly
the same density. The density contrast between the densest components and
the smaller one in each object are $\sim 0.8$ dex for QSO 2350--1241
and $\sim 0.5$ dex for SDSS J0318--0600. 

Next, we determine the ionization parameters characteristic of each 
of the absorption components in both quasars. To this end, we designed
diagnostic plots by which the ionization parameter as well as the total
hydrogen column can be uniquely determined. These plots 
demonstrate that: (1) any given ratio of column densities among
medium and low ionization is a smooth function of the column density 
for a fixed value of
the ionization parameter and (2) these curves of 
column density rations vs. column density are monotonic with $U_H$.
There are various consequences of this: (a) for a given pair of 
measured column densities (and fixed density, chemical composition, and
SED) no more than one solution in $U_H$ and $N_H$
can be found; (b) $U_H$ and $N_H$ and their errors 
are necessarily correlated; (c) these solutions can be found either
graphically or numerically in a way much more efficient than 
though generic numerical optimization techniques; (d) the graphical 
nature of the diagnostic allows one to set the observations from various
different observers on the same page and gain valuable insights.

In the determination of $\vy{U}{H}$ and $\vy{N}{H}$ traditional plots of predicted column densities vs. 
$\vy{N}{H}$  as abscissa have various disadvantages. 
This is because for every column density measured there is whole family of solutions ($\vy{U}{H}$, $\vy{N}{H}$), thus both parameters must be determined simultaneously. A typical approach used is to normalize the predicted column densities to the measured columns and then look for the intersection between various curves. The disadvantages in that is that in dividing theoretical values by measured values mixes up theoretical and observational uncertainties. Further, a solution based on intersections of curves or broad regions, if uncertainties are accounted for in some fashion, offers no simple intuitive understanding of how the results may change in case of systematic effects on either the theoretical modeling or the observations. By contrast, the plots that we propose here clearly allow one to visualize the error bars of the measurements and their significance relative to the predictions of different models. One can also visualize how systematic effects on the calculations, like for example chemical composition or shape of the SED, would shift the results of the diagnostics. Another very important advantage of the proposed plots is the potential to put various kinematic components of the same trough on the same plot and compare them under equal conditions.
It is true that the proposed plots do not show the resulting $\vy{N}{H}$ explicitly, but by fixing 
$\vy{U}{H}$ the whole problem is solved and there is a direct correspondence 
between the observed column density of observed species and $\vy{N}{H}$. Thus, 
$\vy{N}{H}$ can be read directly from the tabulated solutions of the models.

We determined relative distances of the various kinematic components,
firstly under the assumption that all components see the same unattenuated
SED. It becomes immediately clear that the component with the largest column density
is always the first in line from the source.
Once the first kinematic component in line was identified for each
object we include the effect
of attenuation of the SED on the distance determination for the
remaining components. It was found that distance determinations that ignore
attenuation affects are significantly overestimated. By contrast, when
attenuation by the innermost component is considered in the distance 
estimation all the kinematic components in the
absorption troughs are found in close proximity to each other, and possibly related. This result, if found generally true in most FeLoBAL, ought to have
important consequences in our understanding of the dynamics and energetics of 
quasar outflows.

\acknowledgments
We acknowledge support from NSF grant AST 0507772 and from NASA LTSA grant
NAG5-12867.

\end{document}